\documentclass[3p,times,twocolumn]{elsarticle}
 \biboptions{comma,sort&compress}
 
\usepackage{graphicx}
\usepackage{here}
\usepackage{ecrc}

\usepackage{amsmath,stackrel}

\volume{00}

\firstpage{1}

\journalname{Nuclear and Particle Physics Proceedings}

\runauth{}

\jid{nppp}

\jnltitlelogo{Nuclear and Particle Physics Proceedings}

\usepackage{amssymb}
\usepackage[figuresright]{rotating}

\newcommand{\SU}{{\textrm{SU}}}

\newcommand{\vx}{{\mathbf{x}}}
\newcommand{\vp}{{\mathbf{p}}}

\newcommand{\MeV}{\,{\rm MeV}}
\newcommand{\GeV}{\,{\rm GeV}}

\newcommand{\QCD}{{\textrm{\scriptsize QCD}}}

\newcommand{\HRG}{{\textrm{\scriptsize HRG}}}

\newcommand{\cons}{{\textrm{\scriptsize cons}}}
\newcommand{\lat}{{\textrm{\scriptsize lat}}}

\begin{document}

\begin{frontmatter}

%%
%%%%%%%%%%%%%%%%%%%%%%%%%%%%%%%%%%%%%%%%%%%%%%%%%
\title{ Quark-diquark models and baryonic fluctuations in QCD
 $^*$}
 % \corref{cor0}}
 \cortext[cor0]{Talk given at 22th International Conference in Quantum Chromodynamics (QCD 19),  2 - 5 july 2019, Montpellier - FR}
 \author[label1]{E. Meg\'{\i}as\fnref{fn1}}
%  \cortext[cor0]{FAPESP CNPq-Brasil PhD student fellow.}
\ead{emegias@ugr.es}
\address[label1]{Departamento de F\'{\i}sica At\'omica, Molecular y Nuclear and Instituto Carlos I de F\'{\i}sica Te\'orica y Computacional, Universidad de Granada, Avenida de Fuente Nueva s/n, 18071 Granada, Spain}

 \author[label1]{E. Ruiz Arriola}
\ead{earriola@ugr.es}

 \author[label1]{L.~L. Salcedo}
   \fntext[fn1]{Speaker, Corresponding author.}
    \ead{salcedo@ugr.es}

\pagestyle{myheadings}
\markright{ }
\begin{abstract}
We study the baryonic fluctuations of electric charge, baryon number and strangeness, by considering a realization of the Hadron Resonance Gas model in the light flavor sector of QCD. We elaborate on the idea that the susceptibilities can be saturated with excited baryonic states with a quark-diquark structure with a linearly confining interaction identical up to a constant to the quark-antiquark potential, $V_{qD}^\prime(r) = V_{q{\bar q}}^\prime(r)$. We obtain an overall good agreement with the spectrum obtained with other quark models and with lattice data for the fluctuations.
\end{abstract}
% \begin{document}

\end{frontmatter}
%%%%%%%%%%%%
%\vspace*{-1.5cm}

\section{Introduction}
\label{sec:introduction}

In recent years, the thermodynamic approach to strong interactions
pioneered by Hagedorn~\cite{Hagedorn:1965st,Hagedorn:1984hz} where the
vacuum is represented by a non-interacting Hadron Resonance Gas (HRG),
has been used intensively to study the thermodynamics of the confined
phase of Quantum Chromodynamics (QCD). One of the greatest
achievements of this approach has been the recent study of the trace
anomaly, $(\epsilon-3P)/T^4$ with $\epsilon$ energy density and $P$
the pressure. It was computed directly in lattice QCD by several
collaborations~\cite{Borsanyi:2013bia,Bazavov:2014pvz} and within the
HRG approach by using the most recent compilation of the hadronic
states by the Particle Data Group (PDG)~\cite{Tanabashi:2018oca},
leading to an excellent agreement for temperatures below $T \sim 170
\MeV$ (see e.g. Ref.~\cite{Arriola:2014bfa} and references
therein). Thus, this approach has emerged as a practical and viable
path to establish completeness of hadronic states in the hadronic
phase~\cite{RuizArriola:2016qpb,Megias:2017qil,Megias:2016onb}.

There are a number of quark models that try to compute the hadron
spectrum, one of the most fruitful being the Relativized Quark Model
(RQM) for mesons~\cite{Godfrey:1985xj} and
baryons~\cite{Capstick:1986bm}. This model shows that there are {\it
  further states} in the spectrum above some scale as compared to the
PDG, that may be confirmed in the future as hadrons,
although they could also be exotic, glueballs or hybrid states. On the
other hand, the suspicion that the baryonic spectrum can be understood
in terms of quark-diquark degrees of freedom~\cite{Anselmino:1992vg}
has motivated the use of diquark models. Lattice QCD has also provided
insights, as some evidence on diquarks correlations in the
nucleon~\cite{Alexandrou:2006cq} and the dominance of the scalar
diquark channel~\cite{DeGrand:2007vu} have been reported. Previous
studies have traditionally focused on the {\it individual} one-to-one
mapping of resonance states~\cite{Santopinto:2014opa}. In contrast,
the thermodynamic approach allows to perform a more global analysis of
the spectrum, so that possible fine and hyperfine interaction terms are not relevant.

Apart from the equation of state, there are also other thermal observables, like the fluctuations of conserved charges~\cite{Bazavov:2012jq}, that can be used to study the QCD spectrum by distinguishing between different flavor sectors. In the present work, we profit from the new perspective provided by lattice QCD based on the separation of quantum numbers, and try to answer the question whether or not quark-diquark states saturate the baryonic susceptibilities below the deconfinement crossover.

\section{QCD spectrum and thermodynamics}
\label{sec:QCD_spectrum}

\begin{figure*}[htb]
\centering
 \begin{tabular}{c@{\hspace{4.5em}}c}
 \includegraphics[width=0.43\textwidth]{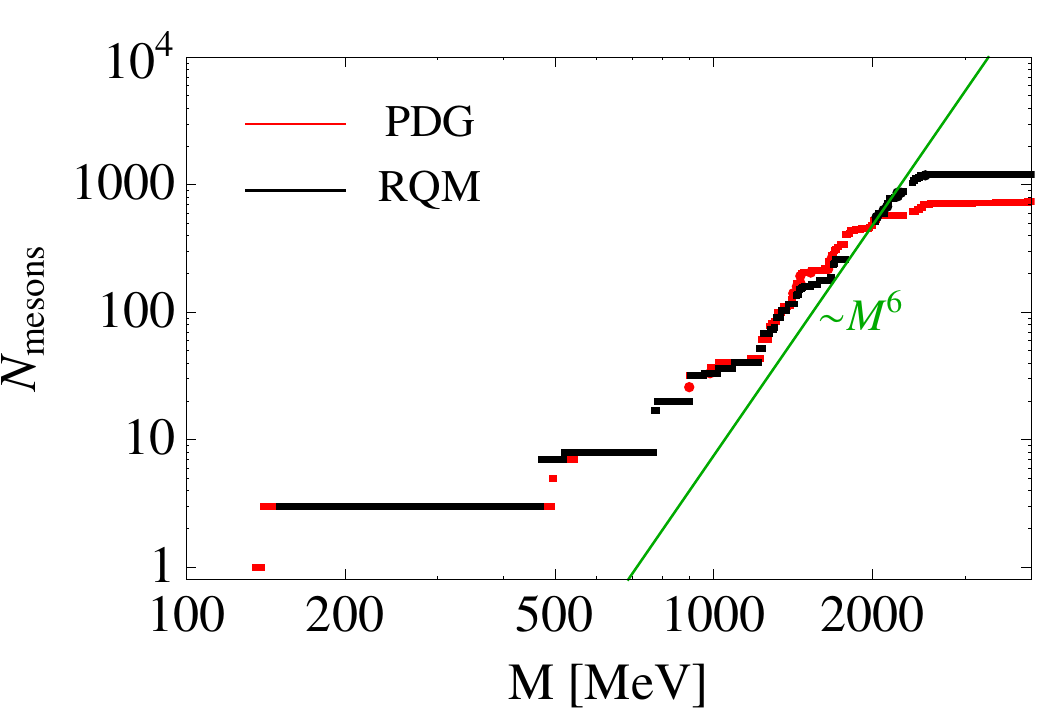} &
\includegraphics[width=0.43\textwidth]{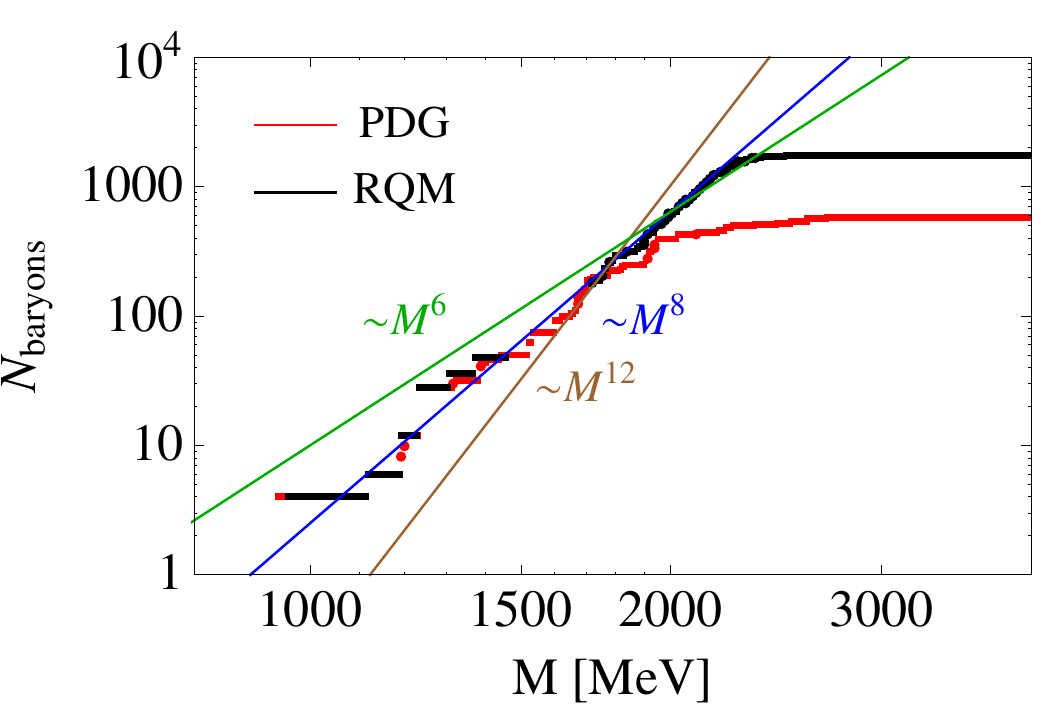} 
\end{tabular}
\vspace{-0.4cm}
 \caption{\it Cumulative numbers for the PDG (red line) and the RQM
   (black line). We display in log-log scale the mesonic states (left panel), and the baryonic states (right panel).  }
\label{fig:nucum-meson-baryon}
\end{figure*}

The cumulative number of states is very useful for the characterization of the QCD spectrum. It is defined as the number of bound states below some mass~$M$, i.e.%
\begin{equation}
N(M) = \sum_i g_i  \, \Theta(M-M_i) \,,
\end{equation}
where $M_i$ is the mass of the i$th$ hadron, $g_i$ is the degeneracy, and $\Theta(x)$ is the step function. So far, the states listed in the PDG echo the standard quark model classification for mesons~$[q{\bar q}]$ and baryons~$[qqq]$. Then, it would be pertinent to consider also the spectrum of the RQM for hadrons, as it corresponds {\it by construction} to a solution of the quantum mechanical problem for both $q \bar q$-mesons and $qqq$-baryons. For color-singlet states, the $n$-parton Hamiltonian takes the form
\begin{equation}
H_n = \sum_{i=1}^n \sqrt{p_i^2+m^2} + \sum_{i< j}^n v_{ij}(r_{ij}) \,,
\end{equation}
where the two-body interactions take the form $v_{q \bar q}(r) = -4\alpha_S/(3r) + \sigma r = (N_c - 1) v_{qq}(r)$. This Hamiltonian for $n=2(3)$ describes the underlying dynamics of mesons(baryons). A~semiclassical expansion of the cumulative number of states~\cite{Caro:1994ht} can be used to study the high mass spectrum for systems where interactions are dominated by linearly rising potentials with a string tension~$\sigma$, in the range $M \gg \sqrt{\sigma}$. At leading order in the expansion, the cumulative number takes the form
\begin{eqnarray}
\hspace{-0.5cm} N_n (M) \!\!\! &\sim& \!\!\! g_n \int \prod_{i=1}^n \frac{d^3 x_i d^3 p_i}{(2\pi)^3} \delta \left({\textstyle \sum_{i=1}^n \vx_i}\right) 
\delta \left({\textstyle \sum_{i=1}^n \vp_i}\right)  \nonumber \\
&\times& \!\!\! \theta (M-H_n(p,x)) \quad \sim \quad   \left(\frac{M^2}{\sigma}\right)^{3 n-3} \,,
\end{eqnarray}
where in the last equality we have neglected the Coulomb term of the potential. Then, one can predict that the large mass expansion of these contributions is 
\begin{equation}
N_{[q{\bar q}]} \sim M^6 \,, \quad N_{[qqq]} \sim M^{12} \,, \quad N_{[q{\bar q}q{\bar q}]} \sim M^{18} \,, \, \cdots 
\end{equation}
We display in Fig.~\ref{fig:nucum-meson-baryon} the separate
contributions of meson and baryon spectra for the PDG and RQM. Note
that while in the meson case the~$N_{[q{\bar q}]} \sim M^6$ behavior
seems to conform with the asymptotic estimate, in the baryon case much
lower powers, $M^{6}-M^8$, than the expected one are identified. We
note that $M^6$ suggests a two-body dynamics, and we take this feature
as a hint that the $qqq$ excited spectrum effectively conforms to a
two body system of particles interacting with a linearly growing
potential. The consequences of this picture will be analyzed in
Sec.~\ref{sec:quark_diquark_model}.

\section{Fluctuations of conserved charges in a thermal medium}
\label{sec:fluctuations}

\begin{figure*}[tbh]
\centering
 \begin{tabular}{c@{\hspace{4.5em}}c@{\hspace{4.5em}}c}
 \includegraphics[width=0.26\textwidth]{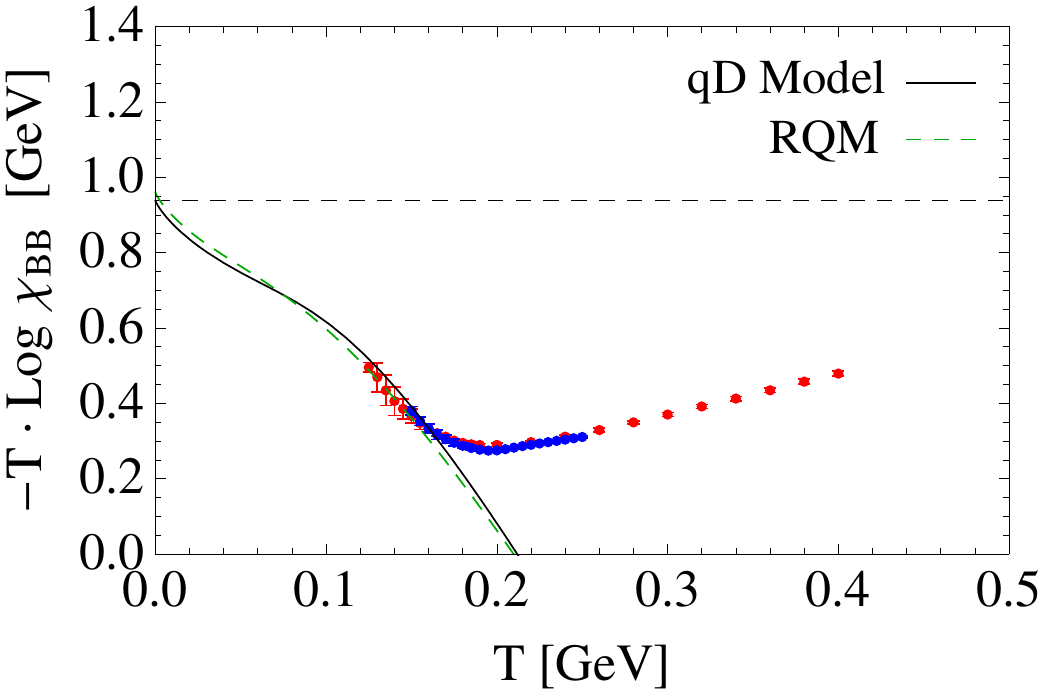} &
 \includegraphics[width=0.26\textwidth]{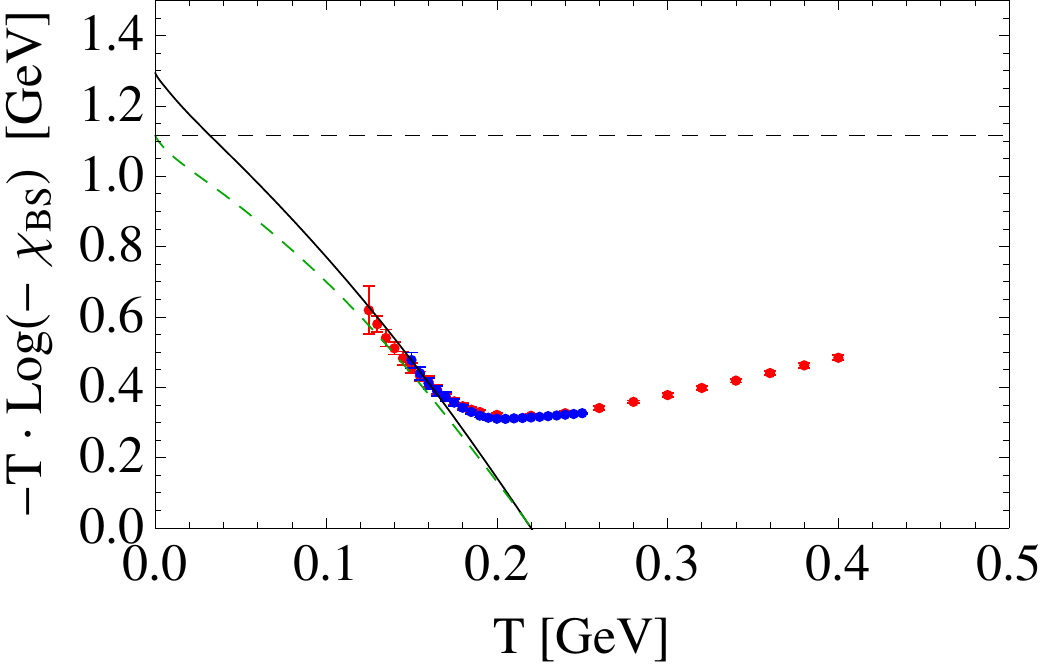} &
 \includegraphics[width=0.26\textwidth]{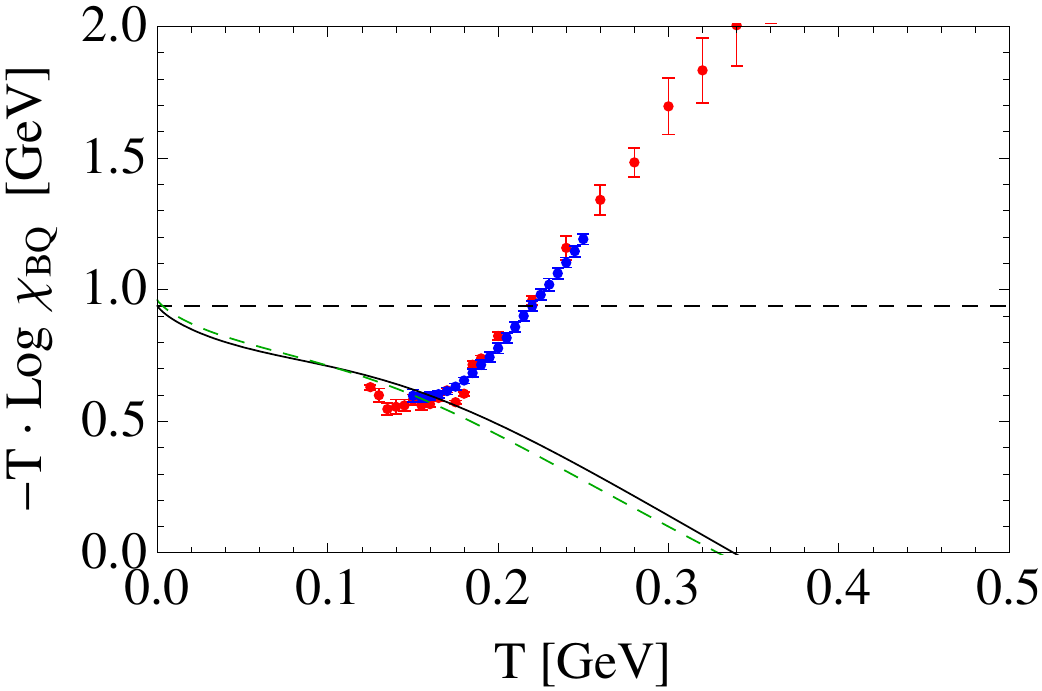} 
\end{tabular}
\vspace{-0.4cm}
 \caption{\it Plot of $-T \log |\chi_{ab}|$ for the baryonic
   susceptibilities as a function of temperature. We display as dots
   the lattice data from Refs.~\cite{Bazavov:2012jq} (blue) and
   \cite{Borsanyi:2011sw} (red). We also display the HRG approach
   results including the spectrum of the RQM~\cite{Capstick:1986bm}
   (dashed green) and the baryon spectrum from the quark-diquark model
   computed in Sec.~\ref{sec:quark_diquark_model} (solid
   black). Horizontal dashed lines represent the values of the
   lowest-lying states contributing to the fluctuations,
   cf. Eq.~(\ref{eq:Chi_lowT}).}
\label{fig:LogChi1}
\end{figure*}

Conserved charges $[Q_a,H]=0$ play a fundamental role in the thermodynamics of QCD. In the ({\it uds}) flavor sector, the conserved charges are the electric charge~$Q$, the baryon number~$B$, and the strangeness~$S$. Their thermal expectation values in the hot vacuum are vanishing, but they present statistical fluctuations that can be computed from the grand-canonical partition function~\cite{Asakawa:2015ybt}
\begin{equation}
Z_\QCD = \textrm{Tr} \exp\left[ - \left( H_\QCD - \sum_a \mu_a Q_a \right)/T \right] \,.
\end{equation}
By differentiation with respect to the chemical potentials, one finds the susceptibilities
\begin{equation}
\chi_{ab}(T) \equiv \frac{1}{V T^3}  \langle \Delta Q_a \Delta Q_b \rangle_T \,, \;\; \Delta Q_a = Q_a - \langle Q_a \rangle_T \,,
\end{equation}
where  $Q_a \in \{ Q, B, S \}$. QCD at high temperature behaves as an ideal gas of quarks and gluons, and in this limit the susceptibilities approach
\begin{equation}
\chi_{ab}(T)  \stackrel[T \to \infty]{\longrightarrow}{} \frac{N_c}{3}\sum_{i=1}^{N_f} q_i^a q_i^b  \,,
\end{equation}
where $q_i^a \in \{ Q_i, B_i, S_i\}$. Within the HRG approach, the charges are carried by various species of hadrons, so that $Q_a = \sum_i q_i^a N_i$, where $N_i$ is the number of hadrons of type $i$. This approach, valid at low enough temperatures, leads to the result~\cite{Megias:2018haz}
\begin{equation}
\begin{split}
\hspace{-0.25cm} \chi_{ab}^{\HRG}(T) =   
 \frac{1}{2\pi^2}  \!\! \sum_{i \in {\rm Hadrons}}
\!\!\!\!\! g_i  \, q_i^a \, q_i^b \sum_{n=1}^\infty  \zeta_i^{n+1} \frac{M_i^2}{T^2} 
 \! K_2\left( \frac{nM_i}{T} \right)  \,, 
\end{split}
\label{eq:chi_HRGM}
\end{equation}
where $\zeta_i = \pm 1$ for bosons and fermions, respectively, and
$K_2(x)$ is the Bessel function of the second kind. This equation
predicts the asymptotic behavior
\begin{equation}
\chi_{ab}(T) \stackbin[T \to 0]{}{\sim} e^{-M_{0}/T}  \,, \label{eq:Chi_lowT}
\end{equation}
where $M_{0}$ is the mass of the lowest-lying state in the spectrum
with quantum numbers $a$ and $b$.  This observation makes it appealing
to plot the lattice data for the fluctuations in a logarithmic
scale. These plots are shown in Fig.~\ref{fig:LogChi1}, and compared
with the HRG approach including the RQM spectrum. We also
display the result by using the spectrum of the quark-diquark model
that is computed in Sec.~\ref{sec:quark_diquark_model}. In the
following we will focus on the baryonic susceptibilities,
i.e. $\chi_{BB}$, $\chi_{BQ}$ and $\chi_{BS}$; as the quark-diquark
picture can only be used to compute the baryon spectrum.

\section{Quark-diquark model for baryons and fluctuations}
\label{sec:quark_diquark_model}

There is nowadays some discussion about the most probable spatial configuration of quarks inside baryons. An interesting possibility would be that the quarks are distributed according to an isosceles triangle, leading to an easily tractable class of models, the so-called relativistic quark-diquark ($qD$) models~\cite{Santopinto:2014opa,Gutierrez:2014qpa,Masjuan:2017fzu}. We will study in this section a simplified version of these models, and use it to compute the baryon spectrum and the baryonic fluctuations.

\subsection{The model}
\label{subsec:model}

In the quark-diquark models, the baryons are assumed to be composed of a constituent quark~$q$, and a constituent diquark~$D \equiv (qq)$. In their relativistic version the Hamiltonian writes~\cite{Santopinto:2014opa}
\begin{equation}
H_{qD} = \sqrt{ \vp^2 + m_q^2 } + \sqrt{ \vp^2 + m_D^2 } + V_{qD}(r) \,. \label{eq:Hamiltonian}
\end{equation}
Using Polyakov loop correlators and Clebsch-Gordan decomposition, we have proven in the static limit that the quark-diquark potential, $V_{qD}(r)$, coincides with the quark-antiquark potential, $V_{q{\bar q}}(r)$, up to an additive constant~\cite{Megias:2013xaa,Megias:2018haz}, a feature which is in marked agreement with recent lattice studies~\cite{Koma:2017hcm}. Hence, we assume
\begin{equation}
V_{qD}(r) = - \frac{\tau}{r} + \sigma r + \mu \,,  \label{eq:VqD}
\end{equation}
with $\tau = \pi/12$ and $\sigma = (0.42\GeV)^2$. The parameters of the kinetic terms in Eq.~(\ref{eq:Hamiltonian}) are controlled by: i) the constituent quark mass, $m_\cons$, and ii) the current quark mass for the strange quark, $\hat{m}_s$; in the following way:
\begin{equation}
m_{u,d} = m_\cons   \,, \qquad m_s = m_\cons + \hat{m}_s \,.  \label{eq:muds}
\end{equation}
In addition, we can distinguish between two kinds of diquarks: scalar $D$, and axial vector $D_{AV}$; and we consider the natural choice
\begin{equation}
m_{D, \textrm{{ns}}} = 2 m_{\cons} \,, \qquad m_{D_{AV}, \textrm{{ns}}} = m_{D, \textrm{{ns}}} + \Delta m_D \,, \label{eq:mDsAV}
\end{equation}
where the subindex $\textrm{{ns}}$ refers to diquarks with nonstrange quarks. Some studies indicate a mass difference between these diquarks of $\Delta m_D \simeq 0.21 \GeV$~\cite{Jaffe:2004ph}, a value that will be adopted in the following. Finally, the breaking of flavor $\SU(3)$ for diquarks will be modeled as
\begin{equation}
m_D = m_{D, \textrm{{ns}}} + n_s  \hat{m}_s  \,, \qquad m_{D_{AV}} = m_D + \Delta m_D \,,
\label{eq:mDsAV2}
\end{equation}
where  $n_s = 0, 1, 2,$ is the number of $s$ quarks in the diquark. With these assumptions, the only free parameters of the model are $m_\cons$, $\hat{m}_s$ and $\mu$. We summarize in Table~\ref{tab:degeneracy} the degeneracies of the states predicted by the model, by distinguishing between their electric charges. For baryonic states,~$B=1$ and $S = -n_s$.

\begin{table}[htb!]
\centering
\begin{tabular}{||c|c|c|c|c||}
\hline\hline
Baryon & $Q= -1$ & $Q=0$ & $Q=1$ & $Q=2$  \\ \hline
$[nn]n$     &  -  &   2  &  2  &  -  \\   \hline
$\{nn\}n$   &  6  &  12  & 12  &  6   \\  \hline
$[nn]s$     &  -  &   2  &  -  &  -   \\  \hline
$\{nn\}s$   &  6  &   6  &  6  &  -   \\  \hline
$[ns]n$     &  2  &   4  &  2  &  -   \\  \hline
$\{ns\}n$   &  6  &  12  &  6  &  -   \\  \hline
$[ns]s$     &  2  &   2  &  -  &  -   \\  \hline
$\{ns\}s$   &  6  &   6  &  -  &  -   \\  \hline
$\{ss\}n$   &  6  &   6  &  -  &  -   \\  \hline
$\{ss\}s$   &  6  &   -  &  -  &  -   \\ 
\hline\hline
\end{tabular}
\caption{Spin-isospin degeneracies of the baryonic states within the
  quark-diquark model. $n$ represents the light flavors~$u,d$. We use
  $[q_1 q_2]$ to denote scalar diquarks, and $\{q_1 q_2\}$ for
  axial-vector diquarks.
  }
\label{tab:degeneracy}
\end{table}

\subsection{Baryon spectrum}
\label{subsec:spectrum}

\begin{figure}[htb]
 \includegraphics[width=0.43\textwidth]{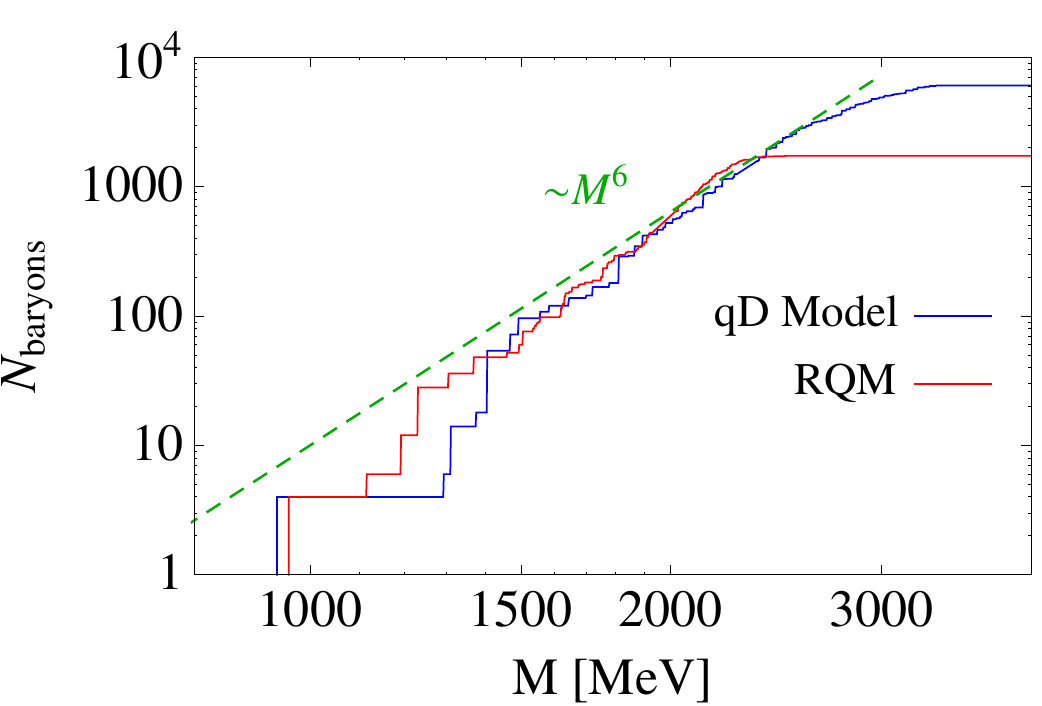} 
\vspace{-0.4cm}
 \caption{\it Cumulative number for the spectrum of baryons as a
   function of the baryon mass. We compare the quark-diquark model of
   Sec.~\ref{sec:quark_diquark_model} with parameters in
   Eq.~(\ref{eq:param}), and the RQM~\cite{Capstick:1986bm}.}
\label{fig:N_baryons}
\end{figure}

The spectrum of the quark-diquark model can be obtained by diagonalizing the Hamiltonian of Eq.~(\ref{eq:Hamiltonian}). The problem does not admit an analytic solution, and we will consider a variational procedure in which the model space is truncated. A convenient basis is that of the 3-dimensional isotropic harmonic oscillator (IHO), which is written in terms of the generalized Laguerre polynomials. The matrix elements of the Hamiltonian are then obtained from 
\begin{eqnarray}
\hspace{-1.3cm} &&\langle n\ell | H_{qD} | n^\prime \ell \rangle =   \int_0^\infty dr \, u_{n\ell}^\ast(r) u_{n^\prime \ell}(r) V_{qD}(r)  \nonumber \\
\hspace{-1.3cm} && + \int_0^\infty dp \, \hat{u}_{n\ell}^\ast(p) \hat{u}_{n^\prime \ell}(p)  \left[ \sqrt{p^2 + m_q^2 } + \sqrt{p^2 + m_D^2}  \right]  \,,
\label{eq:matrix_element_H}
\end{eqnarray}
where $u_{n\ell}(r)$ and $ \hat{u}_{n\ell}(p)$ are the reduced wave functions of the IHO in position and momentum space, respectively~\cite{Megias:2018haz}. A convenient choice of the parameters of the model,
\begin{equation}
\begin{aligned}
&& m_{D, \textrm{{ns}}} = 0.6 \GeV \,, \quad m_{u,d} = 0.3 \GeV \,, \\ 
&&\hat{m}_s = 0.10 \GeV \,, \quad \mu = -0.459 \GeV \,,
\end{aligned} \label{eq:param}
\end{equation}
leads to the spectrum of baryons that is shown in
Fig.~\ref{fig:N_baryons}. We find that below $M < 2400 \MeV$ the
quark-diquark spectrum is in remarkable agreement with the RQM
spectrum where quark-diquark correlations are not assumed a
priori. Let us mention that it is expected that the quark-diquark
picture will be reliable only for excited states, so that in the
following we will use the empirical value of the mass for the nucleon,
$M_n = 938 \MeV$, and apply the quark-diquark model only for the other
baryons.

\subsection{Baryonic susceptibilities}
\label{subsec:baryonic_susceptibilities}

\begin{figure}[htb]
 \includegraphics[width=0.43\textwidth]{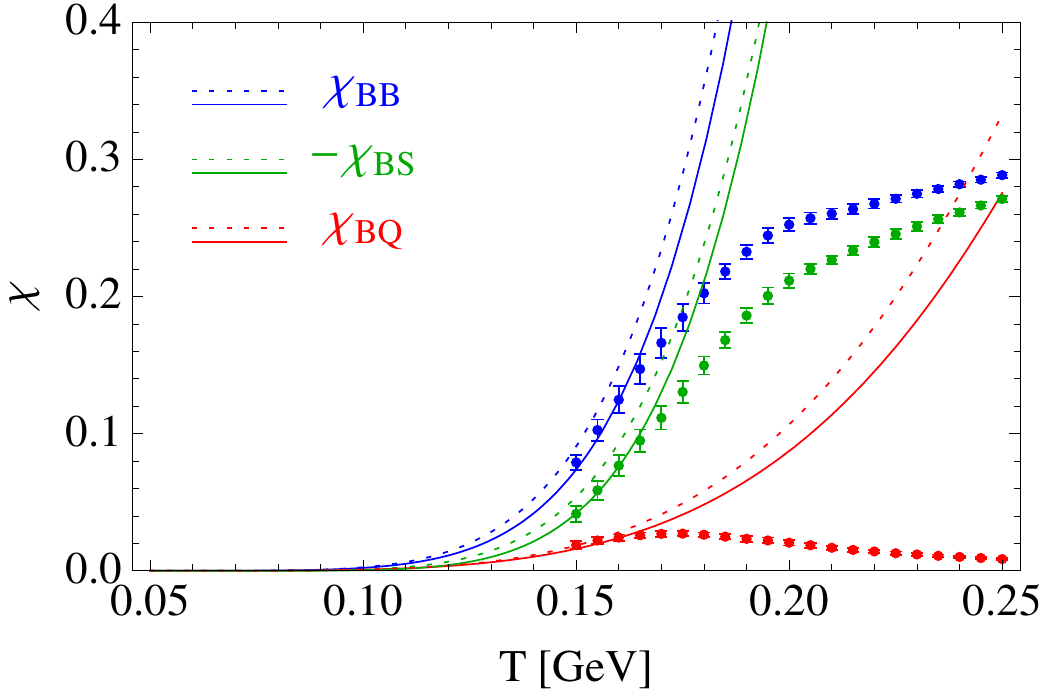} 
\vspace{-0.4cm}
 \caption{\it Baryonic susceptibilities from the quark-diquark model
   (solid) compared to the lattice data of Ref.~\cite{Bazavov:2012jq}.
   We display also as dotted lines the results from the spectrum of the
   RQM~\cite{Capstick:1986bm}. For the quark-diquark model we have
   used the parameters in Eq.~(\ref{eq:param}).}
\label{fig:chi_baryons}
\end{figure}

\begin{figure}[htb]
 \includegraphics[width=0.43\textwidth]{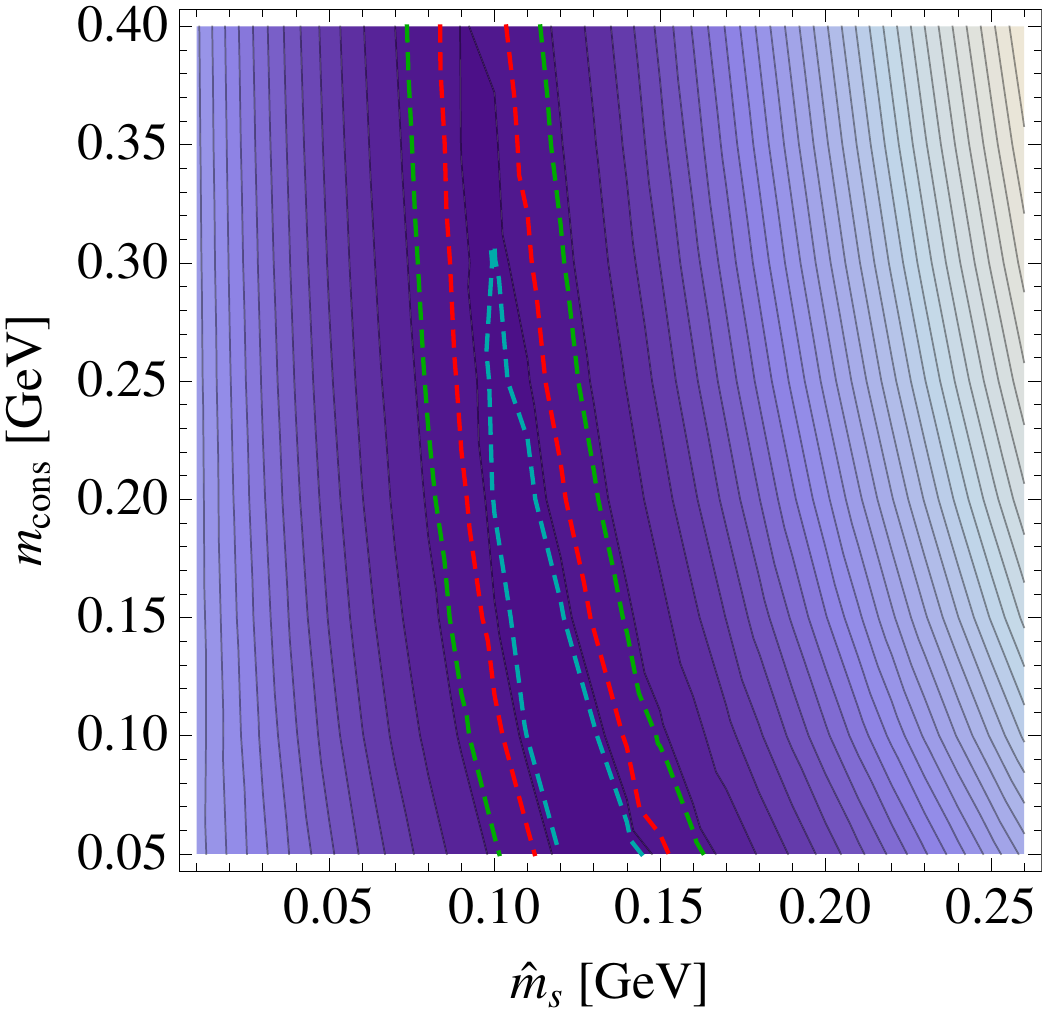} 
\vspace{-0.4cm}
 \caption{\it $\bar\chi^2/\nu$ in the plane $(\hat{m}_s, m_{\cons})$
   from a fit to the lattice data of the baryonic fluctuations
   from~\cite{Bazavov:2012jq} with $T \le 165\MeV$. The dashed lines
   correspond to $\bar\chi^2/\nu = 0.77$ (blue), $\bar\chi^2/\nu = 1$
   (red), and $1+\sqrt{2/\nu}$ (green).}
\label{fig:fit}
\end{figure}

From the spectrum of the quark-diquark model, we can obtain the
baryonic susceptibilities by using the HRG approach given by
Eq.~(\ref{eq:chi_HRGM}). Our goal is to reproduce the lowest
temperature values of the lattice results for these quantities. For
this purpose we have chosen to minimize the function~$\bar\chi^2 =
\bar\chi_{BB}^2 + \bar\chi_{BQ}^2 + \bar\chi_{BS}^2$, where
\begin{equation}
\bar\chi_{ab}^2 = \sum_{j=1}^{j_{\max}} \frac{\left( \chi_{ab}^{\lat}(T_j) -
  \chi_{ab}^\HRG(T_j)\right)^2}{(\Delta \chi_{ab}^{\lat}(T_j))^2}  \,, \label{eq:barchi2}
\end{equation}
and $j_{\max}$ is the number of data points used in the fits. A
typical fit of the model prediction with lattice data leads to the
values of the parameters presented in Eq.~(\ref{eq:param}), and the
corresponding results for the susceptibilities are shown in
Fig.~\ref{fig:chi_baryons}. In order to perform the best fit to the
data, we can study the variation of $\bar\chi^2$ in the full
parameter space of the model. We show in Fig.~\ref{fig:fit} a plot of
$\bar\chi^2/\nu$, where $\nu$ is the number of degrees of freedom, in
the plane $(\hat{m}_s, m_{\cons})$. One can observe that the current
quark mass for the strange quark takes a value compatible with the
PDG, i.e., $80 \MeV \lesssim \hat{m}_s \lesssim 120 \MeV$. In
addition, while the constituent quark mass cannot be determined with
precision, we can ensure that it is in the regime~$100 \MeV \lesssim
m_{\cons} \lesssim 400 \MeV$. Typical values of the parameter $\mu$
entering in Eq.~(\ref{eq:VqD}) are in the range $-0.7 \GeV \lesssim
\mu \lesssim 0 \GeV$.
%%%%%%%%%%%%%%%%%%%%%%%%%
%%%%%%%%%%%%%%%%%%%%%%%%%
%%%%%%%%%%%%%%%%%%%%%%%%%%
%%%%%%%%%%%%%%%%%%%%%%%%%

Some alternative fits have been presented in
Ref.~\cite{Megias:2018haz}. For instance, if the value $\hat{m}_s =
0.10 \GeV$ is adopted and $m_{D, \textrm{{ns}}} = 2 m_{\cons}$ is not
assumed, one finds that the most probable scalar diquark mass is of
the order $m_{D, \textrm{{ns}}} \simeq 0.4 - 0.6 \GeV$, and the
constituent quark mass $m_{\cons} \simeq 0.3 \GeV$. However, these
conclusions depend on the choice of $\hat{m}_s$, leading to lower
values of $m_{D, \textrm{ns}}$ and $m_{\cons}$ when $\hat{m}_s$
increases. By using the spectrum of the quark-diquark model, it has
been studied in Ref.~\cite{Megias:2018haz} some of the baryonic
fluctuations of fourth order as well. While the agreement with lattice
data is reasonable, these data are typically affected by larger error
bars than those of the second-order fluctuations; hence, no firm
conclusions can be extracted from a fit to these~quantities.

%%%%%%%%%%%%%%%%%%%%%%%%%%%%%
\section{Conclusions}
\label{sec:conclusions}
%%%%%%%%%%%%%%%%%%%%%%%%%%%%%

In the present work we have studied the baryonic susceptibilities in a
thermal medium in the $(uds)$ flavor sector of QCD by using the HRG
approach. Being the predictions of this model sensitive to the
spectrum of QCD, this study is also relevant for the characterization
of the spectrum and its completeness. In particular, we have argued
that the asymptotic three-body phase space for confined $qqq$ systems
$\sim M^{12}$ is much larger than the one actually determined in the
RQM, $\sim M^{6}$, which resembles instead a two body systems. This
strongly suggests a dominance of quark-diquark dynamics for excited
baryons, motivating the use of a quark-diquark model to
compute the baryon spectrum. The free model parameters: constituent
quark mass, current quark mass for the strange quark, and an additive
constant entering in the quark-diquark potential; have been determined
from a fit of the baryonic susceptibilities with lattice data. The
results are reasonable and fall in the bulk of previous intensive
studies where a detailed description of the spectrum was pursued. The
extension of this study to nonbaryonic susceptibilities would require
a specific model for mesons which would not be related to the
quark-diquark dynamics. These and other issues will be addressed in a
forthcoming publication~\cite{Megias:2019inprogress}.

%%%%%%%%%%%%%%%%%%%%%%%%%%%%%%%%%%%%%%%%%%%%%%%%
%\vspace*{-0.5cm}
\section*{Acknowledgements} 
%\vspace*{-0.4cm}
%\nin
%%%%%%%%%%%%%%%%%%%%
This work is supported by the Spanish MINECO and European FEDER funds
(Grants No. FIS2014-59386-P and FIS2017-85053-C2-1-P), Junta de
Andaluc\'{\i}a (Grant No. FQM-225), and by the Consejer\'{\i}a de
Conocimiento, Investigaci\'on y Universidad of the Junta de
Andaluc\'{\i}a and European Regional Development Fund (ERDF) (Grant
No. SOMM17/6105/UGR). The research of E.M. is also supported by the
Ram\'on y Cajal Program of the Spanish MINECO (Grant
No. RYC-2016-20678).
%\vspace*{-1cm}
%\vfill \eject
%%%%%%%%%%%

%%%%%%%%%%%
%\section*{References}
%\input{refs}
%\bibliographystyle{elsarticle-num}
%\bibliography{refs}

\begin{thebibliography}{999}


\bibitem{Hagedorn:1965st}
R.~Hagedorn,
\newblock Nuovo Cim. Suppl.~{\bf 3} (1965) 147--186.

\bibitem{Hagedorn:1984hz}
R.~Hagedorn, 
\newblock Lect. Notes Phys. {\bf 221} (1985) 53--76.

\bibitem{Borsanyi:2013bia}
S.~Borsanyi, Z.~Fodor, C.~Hoelbling, S.~D. Katz, S.~Krieg, K.~K. Szabo,
\newblock Phys. Lett. {\bf B730} (2014) 99--104.

\bibitem{Bazavov:2014pvz}
A.~Bazavov, et~al.,
\newblock  Phys. Rev. {\bf D90} (2014) 094503.

\bibitem{Tanabashi:2018oca}
M.~Tanabashi, et~al.,
\newblock  Phys. Rev. {\bf D98}~(3) (2018) 030001.

\bibitem{Arriola:2014bfa}
E.~Ruiz~Arriola, L.~L. Salcedo, E.~Megias,
\newblock Acta Phys. Polon. {\bf B45}~(12) (2014) 2407--2454.

\bibitem{Megias:2016onb} 
E.~Megias, E.~Ruiz Arriola and L.~L.~Salcedo,
\newblock Phys. Rev. {\bf D94}~(9) (2016) 096010.

\bibitem{RuizArriola:2016qpb}
E.~Ruiz~Arriola, W.~Broniowski, E.~Megias, L.~L. Salcedo,
\newblock YSTAR2016 Mini-Proceedings (2016) 136--147, [arXiv:1612.07091].

\bibitem{Megias:2017qil}
E.~Megias, E.~Ruiz~Arriola, L.~L. Salcedo, 
\newblock PoS Hadron2017 (2018) 232.

\bibitem{Godfrey:1985xj}
S.~Godfrey, N.~Isgur, 
\newblock Phys. Rev. {\bf D32} (1985) 189--231.

\bibitem{Capstick:1986bm}
S.~Capstick, N.~Isgur, 
\newblock Phys. Rev. {\bf D34} (1986) 2809.

\bibitem{Anselmino:1992vg}
M.~Anselmino, E.~Predazzi, S.~Ekelin, S.~Fredriksson, D.~B. Lichtenberg,
\newblock Rev. Mod. Phys. {\bf 65} (1993) 1199--1234.

\bibitem{Alexandrou:2006cq}
C.~Alexandrou, P.~de~Forcrand, B.~Lucini,
\newblock Phys. Rev. Lett. {\bf 97} (2006) 222002.

\bibitem{DeGrand:2007vu}
T.~DeGrand, Z.~Liu, S.~Schaefer, 
\newblock Phys. Rev. {\bf D77} (2008) 034505.

\bibitem{Santopinto:2014opa}
E.~Santopinto, J.~Ferretti,
\newblock Phys. Rev. {\bf C92}~(2) (2015) 025202.

\bibitem{Bazavov:2012jq}
A.~Bazavov, et~al., 
\newblock Phys. Rev. {\bf D86} (2012) 034509.

\bibitem{Caro:1994ht}
J.~Caro, E.~Ruiz~Arriola, L.~Salcedo, 
\newblock J.Phys. {\bf G22} (1996) 981--1011.

\bibitem{Asakawa:2015ybt}
M.~Asakawa, M.~Kitazawa,
\newblock Prog. Part. Nucl. Phys. {\bf 90} (2016) 299--342.

\bibitem{Megias:2018haz}
E.~Megias, E.~Ruiz~Arriola, L.~L. Salcedo, 
\newblock Phys. Rev. {\bf D99}~(7) (2019) 074020.

\bibitem{Borsanyi:2011sw}
S.~Borsanyi, Z.~Fodor, S.~D. Katz, S.~Krieg, C.~Ratti, K.~Szabo,
\newblock JHEP {\bf 01} (2012) 138.

\bibitem{Gutierrez:2014qpa}
C.~Gutierrez, M.~De~Sanctis,
\newblock Eur. Phys. J. {\bf A50}~(11) (2014) 169.

\bibitem{Masjuan:2017fzu}
P.~Masjuan, E.~Ruiz~Arriola,
\newblock Phys. Rev. {\bf D96}~(5) (2017) 054006.

\bibitem{Megias:2013xaa}
E.~Megias, E.~Ruiz~Arriola, L.~L. Salcedo,
\newblock Phys. Rev. {\bf D89}~(7) (2014) 076006.

\bibitem{Koma:2017hcm}
Y.~Koma, M.~Koma,
\newblock Phys. Rev. {\bf D95}~(9) (2017) 094513.

\bibitem{Jaffe:2004ph}
R.~L. Jaffe, 
\newblock Phys. Rept. {\bf 409} (2005) 1--45.

\bibitem{Megias:2019inprogress}
E.~Megias, E.~Ruiz~Arriola, L.~L. Salcedo, 
\newblock work in progress (2019).


\end{thebibliography}
%%%%%%%%%%%

\end{document}